\documentstyle{l-aa}

\begin{document}
\input{epsf.sty}

   \thesaurus{05         
              (11.19.4;    
                11.08.1;    
                11.05.2;    
                11.14.1)}   
   \title{"The globular cluster distributions in the Galaxy, M31 and M87: are many 
globulars disappeared to the galactic centres?"
 }

   \subtitle{}

   \author{
          R. Capuzzo--Dolcetta, L. Vignola
           }

   \offprints{R. Capuzzo--Dolcetta}

   \institute{
             Istituto Astronomico, Universit\'a di Roma ``La Sapienza''
             Via G.M. Lancisi 29, 00185 Roma, Italy
             }

   \date{}

   \maketitle

 \markboth{The globular cluster distributions}{Capuzzo--Dolcetta, Vignola}

   \begin{abstract}

The radial distribution of globular clusters in our Galaxy, M31 and M87 is
studied and compared with that of halo stars. The
globular cluster distributions seem significantly flatter
than those of the stars bulge. Assuming this is a consequence of
an evolution of the globular cluster distribution in these galaxies, a
comparison with the (unevolved) stellar distribution allows us to obtain 
 estimates of the number and total mass of clusters lost, which are
possibly gone to feed the massive central objects present in these
 galaxies. 

\noindent
It results that the cluster systems studied should have been initially
about one third and one forth richer than now in our Galaxy and in M31,
respectively, 
and twice as abundant in M87. The estimated mass of globular clusters lost is 
compatible with the galactic nucleus masses.

            
\end{abstract}

\section{Introduction}

The globular cluster systems (GCS) in the two Virgo giant galaxies
M87 and M49 are clearly less concentrated than the halo star
distribution (Harris, 1986). Probably this feature is not common
to all galaxies, even though in many cases the available data are probably not
 good enough to compare reliably the cluster and halo distributions. Anyway,
a safe statement is that no case has been found where the GCS is more 
centrally concentrated than the halo 
(Harris, 1991) 
This is confirmed by the recent HST WFPC2 observation of 14 elliptical galaxies
(Forbes et al 1996).

A possible explanation of this difference in the distributions has
been suggested by Harris \& Racine (1979) and by Racine (1991) as 
a difference in the formation ages of halo stars and globular clusters.
Following these authors, globular clusters are formed earlier, when the
 density distribution 
was less peaked.
This possibility cannot be ruled out, however it is not supported by any
evidence of a significant older age for globular clusters respect to the
halo: this age difference should be large enough to have allowed the
mother galaxy to contract enough to form halo stars in a distribution
as more concentrated than globulars as observed. Note that in disk galaxies
the epoch of cluster formation could be early enough to force chemical
enrichment but not to take on a distinct spatial structure (Harris 1986).
Moreover, this picture does not explain why the tails of the two density
distributions are about the same. 
 Probably, a simpler
explanation, working in the majority of cases, is the coeval birth of globular
 clusters and halo stars with a 
further evolution of the GCS radial distribution, while the collisionless
halo stands almost unchanged. 
The causes of evolution are dynamical friction and tidal interaction with a
compat nucleus; these phenomena can act to deplete the GCS in the denser inner 
galactic
regions so to modify the initial radial distribution just in
the central region leaving unchanged the outer profile, which remains similar
to that of the halo component. If this is true, the halo radial profile
clearly represents the shape of the initial cluster distribution.

\section{Globular cluster system evolution}

There are various indications that the GCS in a galaxy does not
behave like a dissipationless system, as the halo component is. Actually
even if the galaxy is not a spiral where disk shocking is an important 
cause of evolution, the tidal shock due to the passage near to the galaxy
 centre, and dynamical friction (which acts so 
to bring massive clusters closer and closer to the centre) are relevant
causes of GCS evolution. This leads to a more or less important change of
the GCS spatial distribution and mass function. The relevance of the 
mentioned phenomena depends on the galaxy characteristics: triaxiality
enhances the efficiences of both of them (Ostriker, Binney \& Saha, 
 1989; Pesce, Capuzzo--Dolcetta \& Vietri, 1992; Capuzzo--Dolcetta
 1993 (hereafter CD); Capuzzo--Dolcetta 1996). In particular,
Pesce {\it et al.} (1992) showed that clusters on box orbits
in a triaxial potential lose their orbital energy at a rate one order of magnitude
larger than on loop orbits of comparable size and energy
(even if they are quite elongated).
It is very likely that a large fraction, if not all, of globular clusters are 
actually moving
on box orbits and certainly not on quasi--circular loops,
 due to their early formation during the almost radial proto--galaxy
collapse (see Binney 1988).
 So previous evaluations of dynamical friction efficiency 
based on clusters moving on circular orbits were undoubtely 
{\it over--simplified}
and leading to significantly overestimated values of the dynamical
braking time--scales.
This means that massive globulars in triaxial galaxies have probably 
suffered a lot of dynamical braking and have reached the centre of the mother 
galaxy were they can merge to form a super--massive object (not necessarily
a black hole) or can feed a pre--existent one. Of course this
nucleus, if massive enough, can act in a way to shatter incoming globulars
before they are totally orbitally decayed CD
examined in details
the two contemporary effects and found that, assuming as typical globular cluster
masses $10^5$, $10^6$, and $10^7$ M$_\odot$, nuclei as massive as
$5\times 10^6$ M$_\odot$, $2\times 10^8$ M$_\odot$ and $5\times 10^9$ M$_\odot$
are, respectively, needed to effectively halt the infall of globular clusters 
to the potential minimum.

Two equally interesting scenarios
(quantitatively supported in CD) are open:
i) a triaxial galaxy without a primordial massive nucleus can drive the merging 
of a significant mass in the form of orbitally decayed massive globulars 
in a time scale of the order of few $10^8$ yrs,
eventually leading, with modes which are not trivial to be studied, to a
central object massive enough to stop further mass infall; ii) 
a moderately massive primordial nucleus is
 fed by decayed globulars such to produce a gravitational
luminosity in the range of normal AGNs and to grow furtherly
until a steady state is reached. If the mass of the primordial nucleus
is large enough, dynamical friction on globular clusters is overwhelmed by
the tidal shattering, and, moreover, the massive central object also 
changes the orbital structure around it.

\subsection {Is dynamical friction an effective cause of GCS evolution ?}

The actual role of dynamical friction in galaxies 
has been questioned on the basis
of various arguments which actually apply just to CGSs
in M31 and M87, two of the best studied cases.
It is worth nothing that even if it is plausible that its role is important
whenever clusters move on almost radial orbits, the quantitative definition
 of the relevance
of dynamical friction and tidal destruction has been done just in a triaxial galaxy
(Pesce et al, 1992; CD; Ostriker et al, 1989). 

Now,
while it is now quite accepted that the inner part of M31 is triaxial
(Lindblad 1956; Stark 1977; Bertola {\it et al.} 1991) and so it is a galaxy
where dynamical friction and tidal distruption effects should be 
significantly enhanced, a triaxiality for M87 is not evident.
Good CCD photometric data by Zeilinger {\it et al.} (1993) for the inner
M87 region show almost round isophotes in the inner M87 region ($r < 3''$) and
a twisting occurs at $3 ''$ from the centre, the major axis being shifted to
a position perpendicular to the jet and the ellipticity grows up to $0.2$
at $r \simeq 80''$.
This means that M87 is not necessarily one of the best candidate to investigate
about the evolution of CGS distribution.

\noindent
With regard to M31 two serious observational points are:

\noindent
i) the M31 galactic nucleus seems to be significantly redder than 
            globular clusters (Surdin and Charikov, 1977),
\noindent
ii) M31 globular clusters seem to show a trend of increasing metallicity toward 
the galactic centre (Huchra et al 1991),
anyway this trend , see Fig. 2 in van den 
Bergh (1991) is quantitatively questionable.

Let us explain why in our opinion points i) and ii) are much less serious
indication against the importance of dynamical friction and GCS evolution 
mechanism than it is superficially thought.

First of all the above mentioned data do not constitute
a significant sample to extract general conclusions, referring 
just to one galaxy, anyway a trend of
redder integrated colours towards the centres is a common feature
of many galaxies (Gallagher {\it et al.} 1980),
 and needs in any case
an explanation.

\noindent 
According to various authors, due to the apparent high metallicity of central
region of M31, the decayed 
high mass
clusters should have been more metal abundant (redder) than the only presently
observed, and this needs a correlation between mass and metallicity for
globular clusters. 

\noindent
This metal abundance--mass correlation for globular clusters is claimed to
be {\it ad hoc} because of the poor correlation presently observed 
  between metal content and total {\it luminosity} of galactic globulars.
There is an important caveat before concluding that also a $Z$
 mass--correlation 
is not holding. It comes by the fact that the $Z$--L correlation cannot be
 considered 
exactly representative of
a $Z$ mass--correlation because the mass--luminosity ratio actually
depends on the metal content, and it increases with $Z$.

\noindent
This means that a
flat $Z$--L correlation transforms into a (more or less steep) {\it increasing}
$M(Z)$. Moreover, what is observed now is the {\it present} luminosity--$Z$
correlation, which, in the case of efficiency of the dynamical friction
braking is biased (respect to the initial) towards lower luminosities (masses)
and
metallicities,
likely hiding an initial stronger correlation.

A mass--metallicity relation for globular clusters is, anyway,
not merely an {\it ad hoc}
hypothesis, being verified for instance, 
in galaxies: 
the brighter galaxies contain redder, in the average, globulars (see
van der Bergh 1991 for M31 and our Galaxy's clusters)
and has a physical
interpretation on the basis of a steeper, with increasing mass, potential well to
be overcome by the enriched material expelled by SNs. 

\noindent
Another point that seems hardly compatible with the claimed efficiency
of dynamical friction (acting more on massive globulars) is that in M87
no dependence of the globular cluster
luminosity function on galactocentric distance is found.

\subsection{The radial dependence of the GCS luminosity function}

There are various reasons why the lack of evidence of a radial dependence of 
GCS luminosity function
is not a significative point against dynamical friction to
be occurred on clusters:
i) even if a spatial trend of the luminosity function is present (massive
 globulars moved to inner regions) it is expected to occur in quite
central regions (within the bulge star core radius, see Capuzzo--Dolcetta
 and Tesseri 1996) not easily covered by proper observation;
ii) projection effects weaken any radial trend of the luminosity
function;
iii) dynamical friction
reduces the average galactocentric distance of massive GCs more than that of
light globulars, but this contemporarily means they are shifted to inner
galactic zones where they likely lose their individuality because they become
hardly observable and more easily destroyed by the intense tidal field.

Let us give some quantitative support to point ii).
Suppose to have a sample of globular clusters whose mean mass $<m>$ varies
with the galactocentric distance in a way to have smaller masses in external
regions ($<m>$ varies from $10^6$ M$_\odot$ to $10^5$ M$ \odot$ going from
the centre to 5 times the core radius) and with a mass 
spectrum corresponding to
a gaussian V-magnitude function characterized by a dispersion
around the mean magnitude ($(M/L)_{V\odot}=1.6$ is assumed) which is larger
($\sigma_V^2=3$)
in peripheral galactic regions than around the centre ($\sigma_V^2=1.5$)
(this is what qualitatively expected when 
dynamical friction and tidal disruption have been effective).  
Assuming the GCS distributed spherically according to the modified Hubble
profile
\begin{equation}
n(r)= n_0 \left[{1+\left({r \over r_c }\right)^2 }\right]^- {3 \over 2}
\end{equation}
we can compare the volume LF with the projected LF, sampled at various 
distances from the centre in a galaxy with a distance modulus $(m-M)_V = 31.3$
(similar to Virgo cluster) (see Fig. 1 a,b).

\noindent
Projection should reduce the difference among the peak magnitudes and 
the widths of the LF sampled at various galactocentric distances.
 Actually, Fig. 1 
shows how a V-peak difference of 4 mag 
reduces to just
2.5 mag, while the width of the projected LF results almost constant 
(independent on the galactocentric distance). 

\noindent
We conclude that 
a radial dependence of the width of such LFs would be undetectable,
while the detection of a variation in the V-peak would require 
globular cluster sampling {\it well within} the galaxy core,
because (as it is seen in Fig. 1 b) the V-peaks of the LF sampled 
at $r=r_c$ and $r=5r_c$ differs for a quantity similar to the standard deviation
of the mean ($\sigma_\mu \simeq 0.1~mag$ for a typical total sample of
$\approx 500$ clusters). 
This explains why, even if evolutionary effects have been active on GCSs,
LF radial trends have not been detected in the past. Higher resolution
observations to have larger sample abundances in inner galactic regions
are needed.
\begin{figure}
\epsfxsize=8.truecm
\epsfbox{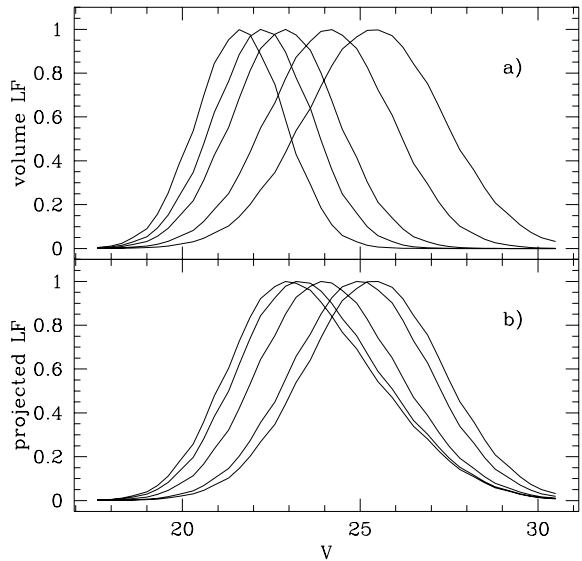}
 \caption{ panel a: normalized V-magnitude volume luminosity 
functions of the GCS at various galactocentric distances 
($r/r_c=0, 0.25, 0.5, 1, 5$);
  panel b: projected luminosity functions at the same (projected)
  galactocentric distances as in panel a.}
   \label{Fig.1}
  \end{figure}

\section{Present and initial radial distributions}

A way to estimate the number of globular clusters lost
 during the evolution of a globular
cluster system was suggested by McLaughlin (1995) who applied it to the
M87 galaxy.
This estimate is based on 
the assumption that the stellar bulge and the
 globular cluster distributions were initially the same (due to a co-eval
 formation) and on that the shape of the stellar
distribution has remained unchanged. The first hypothesis is
supported by the observed
similiarity of the stellar bulge and globular cluster projected
profiles in the outer regions
(outside a certain distance $\bar{r}$ from the centre)
of various galaxies. 

\noindent
The second hypothesis stands firmly on that the bulge is a collisionless
system.
The initial globular cluster distribution $n_0(r)$
 (assumed to be equal in shape to the
present stellar distribution) is, practically, obtained by a scaling of the
 present GCS distribution to the stellar one.
Of course also the projected initial density profiles, $\sigma_0(r)$,
are assumed to be the same.
Once $n_0(r)$ or $\sigma_0(r)$ have been determined, the number of missing 
clusters is given by the integral over the whole galaxy of the difference between
 $n_0$ and $n$ (or $\sigma_0$ and $\sigma$). 

\subsection{The data sets and interpolations}

We consider the well established data sets of CGS in
our Galaxy, M31 and M87, this latter mainly for the
 sake of a comparison with McLaughlin's
 (1995) results.

\noindent
To fit the stellar distribution of M31 and M87 we used the
model of de Vaucouleurs(1958) and for our Galaxy the Young's model (1976); 
good fits to the globular cluster distributions are obtained by means of
the empirical King's models (King 1962).

\noindent
 To obtain the initial 
globular cluster distribution we vertically shift the stellar distribution
to match, in the external region ($r \geq \bar{r}$), the GCS distribution,
so to have a scaling factor, 
$d$, depending on $\bar r$. Thus, we can represent
the initial globular cluster distribution by:

  \begin{equation}
   \log n_0=\log n_s + d(\bar{r})
   \end{equation}

  \begin{equation}
   \log \sigma_0=\log \sigma_s + d(\bar{r})
   \end{equation}

\noindent
were $n_s(r)$ and $\sigma_s(r)$ are the (observed) stellar distribution
volume and surface densities.
Hereafter the volume and surface number densities will be given in 
$kpc^{-3}$ and $kpc^{-2}$ respectively.
\section{Number of clusters missing in our Galaxy and M31}

\subsection{Our Galaxy}
The most complete set of data of galactic globular clusters is still
 given by
Webbink (1985). It refers to 
154 globular clusters. For our purposes we need only  
the distance of each cluster from the galactic centre.
A good fit to the present distribution is:

     \begin{equation}
     n(r)=\left \{\left[\frac{1}{1+\left(\frac{r}{2}\right)^2}
     \right]^{0.5}-
     \left[\frac{1}{1+\left(\frac{b}{2}\right)^2}\right]^{0.5} \right \}^3
     \end{equation}
where $b=49$.

\noindent
As distribution of bulge stars in our Galaxy we use the Young's model (1960)
(see Fig. 2).
Thus we obtain the initial globular cluster distribution:

     \begin{equation}
     n_0(r)=426 exp\left[-7.669 \cdot \left(\frac{r}{2.7}\right)^
      {0.25}\right]
      \left(\frac{r}{2.7}\right)^{-0.875}
     \end{equation}

The estimated number of globular clusters lost for our Galaxy is
$N_l=56$, i.e. about $36\%$ of the present sample's abundance.
\begin{figure}
\epsfxsize=8.truecm
\epsfbox{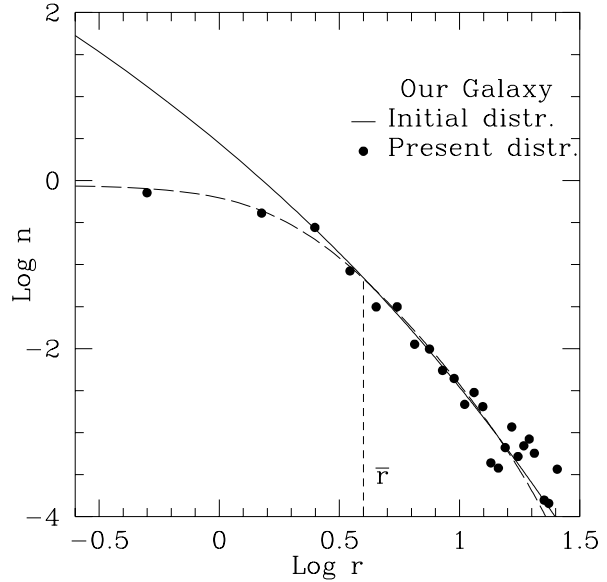}
             \caption{The globular cluster initial distribution
             and the present one
               for our Galaxy. The dashed line refers to the point where 
             the vertically shape bulge and GCS profile start to overlap}
       \label{Fig.2}
    \end{figure}

\subsection{Andromeda (M31)}

Various compilations of data for globular clusters in M31
are available. We refer to the "Adopted Best Sample" of Battistini
 {\it et al.} (1993) compilation,
for it is the most complete and best discussed source of 
data for the radial distribution of globular clusters in this galaxy.

\noindent
The flattening of the M31 globular cluster distribution compared to that
of the star spheroidal component was first noted by de Vaucouleurs \& Buta 
(1978), later questioned by Wirth, Smarr \& Bruno (1985) who claimed a 
large part of this flattening as being due to incompleteness.
The completeness of the Battistini
{\it et al.} data in the inner bulge region is well addressed, and 
a residual flattening of their adopted samples respect to the spheroidal 
star component is evident.

Matching the globular clusters data with
a King's model leads to the following analitycal fit to the present globular cluster
distribution:

\begin{equation}
\sigma(r) = 16.7 \left
    \{\left[\frac{1}{1+\left(\frac{r}{0.7} \right)^2}\
    \right]^{0.5}-
    \left[\frac{1}{1+\left(\frac{a}{0.7}\right)^2}\right]^{0.5} 
    \right\}^{1.52} 
\end{equation}
were $a=30$.
The data for the stellar distribution are taken from de Vaucouleurs (1958),
taking the luminosity distribution as representative of the stellar
density distribution (see Fig. 3). We obtain the initial
 globular clusters distribution:

   \begin{equation}
    \sigma_0(r) = 785.2 \cdot exp \left[-7.427 \cdot r^{0.2} +3.1\right]
    \end{equation}

\begin{figure}
\epsfxsize=8 truecm
\epsfbox{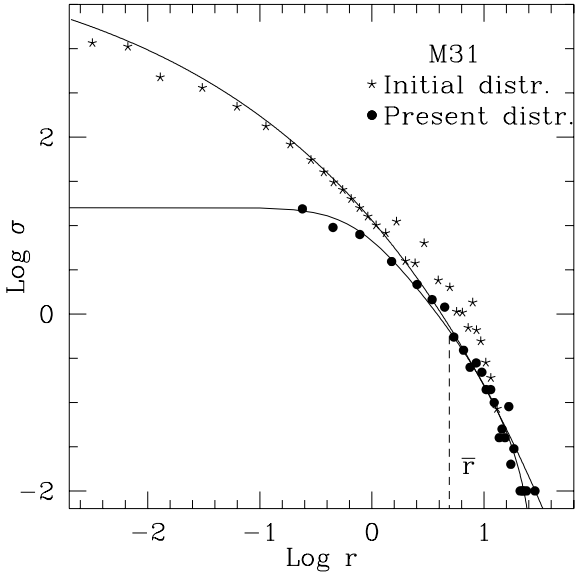}
             \caption{Initial  and present globular cluster distributions
              in M31}
       \label{Fig.1}
    \end{figure} 

The surface integral of $\sigma_0(r)-\sigma(r)$ gives
$N_l \simeq 76$ as number
of globular clusters lost; this is $25 \%$ of the present number.

\subsection{Evaluation of globular cluster mass fallen to
the galactic centres}

An approximate value of the mass fallen to the centre 
of
M31 and our Galaxy can be given by mean of the knowledge of $N_l$ 
and of the average mass of destroyed globular cluster $<m_l>$.
The determination of $<m_l>$ requires a detailed evaluation of the
 tidal distruption and dynamical friction effects on an assumed initial mass
 function.
By the way, due to that the two phenomena erode the CGS
on apposite sides of the mass function, the mean value of the globular cluster mass $<m>$
is not expected to change very much in time whenever the initial mass 
function is not too asymmetric, and thus it can be chosen as a good reference value
for $<m_l>$. This is confirmed by results obtained with a
 theorical model (Capuzzo--Dolcetta \& Tesseri 1996) under the hypotesis of a
constant initial initial mass function ($\Phi(m)~=~constant$).

The knowledge of
the mean mass of globular clusters,  $<m>=3.2 \cdot 10^5 M_\odot$ for
our Galaxy and $<m>=2.7 \cdot 10^5 M_\odot$
 for M31,  gives as mass lost $M_l = 1.8 \cdot 10^7 M_\odot$
 and $M_l = 2.1 \cdot 10^7 M_\odot$ respectively.
These values should be compared with the nucleus masses in our galaxy
($3 \cdot 10^6 M_\odot$, see Krabbe et al 1995) and in M31 
($10^7 M_\odot$, see e.g. Melia 1992).

\subsection{Sources of error}
 
A significant source of error evaluation of $N_l$ is the
 indetermination in the of the region where the globular 
cluster distribution profile is evolved, i.e. the estimate of
 $\bar{r}$.
A relative error in $\bar{r}$ induces an error in $N_l(\bar{r})$:
\begin{equation}
   \frac{\Delta N_l}{N_l}= \frac{\partial N_l}{\partial  \bar r}
   \frac{\bar{r}}{N_l} \frac{\Delta \bar{r}}{\bar{r}},
\end{equation} 

\noindent
being:

\begin{equation}
\frac{\partial N_l}{\partial \bar r}= \ln~10 \cdot 2 \pi \cdot 
10^{d(\bar r)} \cdot d'(\bar r)~
\int_0^{\bar r} \sigma_s(r)  r~ dr
\end{equation}
or, when the spatial density is available:
\begin{equation}
\frac{\partial N_l}{\partial \bar r}=\ln~10 \cdot 4 \pi  \cdot 
10^{d(\bar r)} \cdot d'(\bar r)~
\int_0^{\bar r}
n_s(r) r^2 ~ dr
\end{equation}
where $d'(\bar r)$ is the derivative of $d(\bar r)$ with respect to $\bar r$.

\noindent
For M31 and our Galaxy we find that the error
${\Delta {\bar r}}/{\bar r}$
reflects in relative errors ${\Delta N_l}/{N_l}$ given by 
$0.75 {\Delta {\bar r}}/{\bar r}$ and  
$0.63 {\Delta {\bar r}}/{\bar r}$, respectively.

\section{M87}

\noindent
For the sake of comparison with previous work (McLaughlin 1995) 
we applied our method to M87.
The data are
taken from McLaughlin (1995) and from de Vaucouleurs
 and Nieto (1978, 1979) for globular clusters and halo stars, respectively.
The fits we obtained from those distributions are:
 
   \begin{equation}
   \sigma(r)=15.5 \left \{\left[\frac{1}{1+\left(\frac{r}{1.6}\right)^2}
    \right]^{0.5}-
    \left[\frac{1}{1+\left(\frac{c}{1.6}\right)^2}\right]^{0.5}\right \}
    \end{equation}
 
   \begin{equation}
   \sigma_0(r)=67.62 \cdot exp \left[-3.848\left(\frac{r}{1.543}\right)^{0.349}+
    3.875\right]
    \end{equation}
where $c=60$.

We have fixed $\bar r$ as
the point where the two distributions clearly show the same shape,
as it is shown in Fig. 4 ($\bar r= 12.5 kpc$).

The number of globular clusters lost 
is found to be $N_l = 4018$, which is about the same number of globular
clusters presently observed.

We have numerically evaluated the error induced on $N_l$ by an error in
{$\bar r$}, finding that 
an error ${\Delta {\bar r}}/{\bar r}$ reflects in a relative
errors for ${\Delta N_l}/{N_l}\simeq 2.12
{\Delta {\bar r}}/{\bar r}$; i.e. much greater than for our Galaxy and 
M31.

\begin{figure}
\epsfxsize=8.truecm
\epsfbox{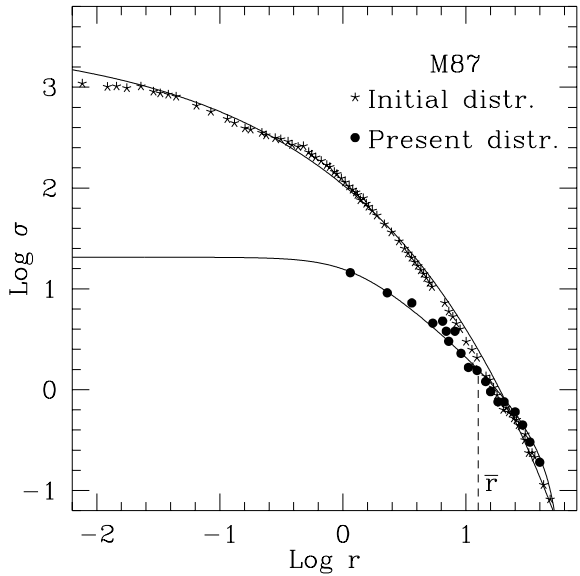}
            \caption{The initial and the present globular cluster 
                  distribution in M87
               }
       \label{Fig.3}
    \end{figure}
This large sensitivity of $N_l$ on the choice of $\bar r$ is part of the 
explanation of the great difference between our value of $N_l$ and
that given by McLaughlin ($N_l \simeq 1150$).
This difference is indeed accounted by:

i) we have considered as best value
for the radius $\bar r \simeq 12.5~ kpc$ instead of $8~kpc$ (used by
McLaughlin 1995): it seems that the two 
distributions have a very similar shape only outside $12.5~ kpc$.
The value of $8~~ kpc$ for $\bar r$ seems an underestimate. Infact
as shown by Fig. 5, the distribution obtained shifting vertically the bulge
star profile to intersect the present GCS distribution at $\bar r=8 kpc$ 
represents an acceptable initial distribution for the GCS just if we 
accept as realistic that in external regions($r \geq 8 kpc$) there are
at present more clusters than initially.
This implies the existence of a mechanism which populated the external regions,
while dynamical friction and tidal distruption depopulated the inner regions.
 
ii) the analytical fits are obtained with different functions: McLaughlin
used, for both globular clusters and star bulge isotropic, single mass
King's (1966) models,
while we used an empirical
King's (1963) model to fit the present globular cluster distribution and the
de Vaucouleur's (1958) model (more peaked than the King's 1966 one) to fit the
bulge distribution.
Anyway the difference in the analytical fits can account for just a $30 \%$
difference in $N_l$. Infact, we obtain $N_l\simeq 1590$ instead of
$N_l\simeq 1150$ integrating our distribution up to the same radius ($\bar r
=8 ~kpc$) used by McLaughlin.

Now, if we take as mean mass of the globular clusters in M87 
$<m> = 6.6 \cdot 10^5 M_\odot$ (McLaughlin 1995) our estimate of 
the mass lost is
$2.65 \cdot 10^9 M_\odot$. If we compare this value with the nucleus mass of
M87 (which is estimated to be $ \simeq 2.4\cdot 10^9 M_\odot$
within $18 pc $ of the nucleus, see Ford and al 1994) we see,
also in this case,
that the distruption of globular clusters could have strongly influenced the
formation of the
central nucleus of this galaxy.

\begin{figure}
\epsfxsize=8.truecm
\epsfbox{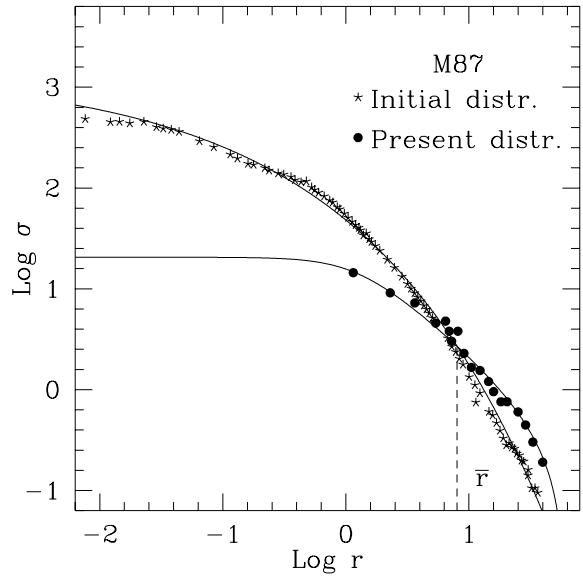}
            \caption{As in Fig. 3, but with $\bar r = 8~kpc$
               }
       \label{Fig.4}
    \end{figure}

\section{Conclusions}

It is both a reasonable and simple
hypothesis that the globular cluster and spheroidal
components of galaxies formed contemporarily during the first stages
of protogalaxy collapse so to have, initially, the same spatial distribution.
Observed (and kinematic) spatial differences 
should be explained on the basis of evolution of the globular cluster 
system (GCS). 

We have given reference to quantitative studies which 
point out the
role of dynamical causes of this evolution in galaxies where
clusters move on sufficiently radial orbits. We have also explained
why most of the observational data available is, at present,
at all unsufficient to rule out that such an evolution occurred. 
To state something meaningful, observations of clusters in
the innermost regions (i.e. within the bulge core)
 to compare with clusters in outer regions of their parent galaxy,
 as well as kinematic data to determine the cluster velocity ellipsoid
are needed.
\par
Through the comparison between 
the globular cluster and spheroidal radial distributions we
determined the number, $N_l$, of clusters lost in our Galaxy, 
M31 and M87.
We found that the GCSs of our Galaxy, M31 and M87 should have been initially
1.45, 1.25, and 2 times more popolous than now.

\noindent
The mass of missing 
clusters has likely gone to the centre of the parent galaxy, where it can
contributed to enrich the nucleus by an amount of the order of
$N_l <m>$, where $<m>$ is the (present) mean value of cluster mass.
This corresponds to $1.8\cdot 10^7M_\odot$, $2.1\cdot 10^7M_\odot$
and $2.7\cdot 10^9M_\odot$ for the Galaxy, M31 and M87, respectively.
It is relevant noting that these values are all very similar to 
available estimates of the nucleus masses in these galaxies.

\end{document}